# Variational design for a structural family of CAD models

Qiang Zou, Qiqiang Zheng, Zhihong Tang, Shuming Gao*

State Key Lab of CAD&CG, Zhejiang University, Hangzhou, 310027, China

**Abstract**

Variational design is a well-recognized CAD technique due to the increased design efficiency. It often presents as a parametric family of solid models. Although effective, this way of working cannot handle design requirements that go beyond parametric changes. Such design requirements are not uncommon today due to the increasing popularity of product customization. In particular, there is often a need for designing a new model out of an existing structural family of models, which share a structural pattern but have individually varied detail features. To facilitate such design requirements, a new method is presented in this paper. The idea is to express the underlying structural pattern in terms of a submodel composed of the maximum common design features of the family, and then to build a single master model by attaching to the submodel all detail design features in the family. This master model is a representative model for the family and contains all the features. By removing unwanted detail features and adding new features, the master model can be easily adapted into a new design, while keeping aligned with the family, structurally. Effectiveness of this method has been validated by a series of case studies and comparisons of increasing complexity.

**1. Introduction**

Computer-aided design (CAD) is an important industrial practice used in many applications, including automotive, shipbuilding, and aerospace. With the help of CAD tools, a large majority of part models are not constructed from scratch today but are modified versions of existing models, leading to what is commonly termed as variational design (or adaptive deign) [1–3]. This CAD modeling means can reduce design time significantly. Given the competitive pressures in today's market, variational design can make an important contribution towards shorter time-to-market and lower labor costs.

Variational design is often carried out using a solid modeler, a field pioneered by Herb Voelcker [4–6]. Design variations are generated by varying parameters pre-defined in solid models [7]. Parametric solid modeling has proven effective in many engineering tasks, especially for those involving exact design requirements. It, however, restricts design variations to a parametric family of parts. This limitation manifests itself through dynamic design requirements that ask for, for example, adding new features and/or removing existing features [8]. Dynamic design requirements are not uncommon today due to the increasing popularity of product customization [9].

Although going beyond parametric variations, the desired new design is not likely to deviate from the design to be reused too much, otherwise we lose the advantages of variational design. Particularly, they often share main functions but have individually customized detail functions. There is thus a shared structural pattern between their CAD models. One typical example is the rotor discs of a jet engine; they are similar in their functions among frames but have different dovetail slots (i.e., detail functions) [10,11]. As product customization goes on, a company will accumulate a collection of structurally similar CAD models (to be called a structural family hereafter). The problem then arises: given a structural family of CAD models, how to easily and quickly carry out variational design when a new product customization order comes in.

The above problem involves two essential tasks: (1) retrieving a base model fittest for the new purpose; and (2) modifying this model according to the new purpose. Without proper computer assistance, the user needs to manually thread through all models in the family to find the base model. Tedious model comparisons may also be needed to pick out detail features distributed in individual models and useful for the new design. Combining those detail features to the base model is, again, not trivial because of complex feature dependencies. To makes the situation even worse, such tedious manual work must be repeated whenever a new design is carried out. (Section 3 will give a more detailed discussion about these limitations.)

This paper presents our attempts to address the above problem through reducing the manual effort involved. The idea is to construct automatically a master model for the family, which can be quickly modified to accommodate required design changes. The master model is a representative model for the

---

family and contains all features. With it, all models in the family can be expressed in terms of removing unwanted detail features from the master model. New models can be generated in a similar manner, possibly with an additional operation of adding new detail features. Clearly, models generated in this way keep aligned with the family, structurally. Then, variational design for a structural family is made easy and quick.

The master model is not a simple duplication and then union of all features in the family, but another feature model retaining not only all features but also their dependencies in individual models. This work solves this problem by formulating it as a maximum submodel extraction problem. This submodel is composed of the maximum common features among all models in the family. With this submodel, the intended master model is generated by attaching all remaining features to it, with regards to their feature dependencies.

The rest of the paper is organized as follows. Section 2 introduces the related work. Section 3 gives a brief overview of our approach. Section 4 describes the specific algorithms used to generate the master model. Section 5 presents variational design examples based on the proposed master-based method. Section 6 concludes this paper.

**2. Related work**

Many researchers have investigated the importance of variational design [7,12–17]. Industrial reports also confirmed its advantages—up to 80% reduction of design time has been seen [18]. This practical significance has motivated many research studies in the last decades. Although presented in various forms, central to them are two tasks: finding a CAD model to reuse, and modifying the model. If variational design is done *a priori,* such as parametric modeling (i.e., modeling of part families), there is no need for the former task. It is only necessary when variational design is done *a posteriori*, as in common design structure recognition. In this regard, methods related to this paper are categorized into priori and posteriori, and their effectiveness are discussed with regards to the above two tasks, especially for handling a structural family of parts.

**2.1. A priori variational design**

Priori means that a design's possible variations are taken into consideration when modeling it with a CAD tool. Because all major CAD modelers are based on solid modeling, the essence here is to incorporate editable definitions into solid models [1]. The first widely accepted method of doing so is parametric modeling. It uses parameterized features to construct a solid model, and those parameters are the editable definitions [19,20]. The solid model's actual shape is a function of parameters, and varying them instantiates a family of models [21]. As such, a parametric model itself is the base model for variational design. There is no need to do model finding (i.e., the previously noted first task).

Despite the success in many applications, parametric modeling has some important limitations, as documented in [7,22–26]. Particularly, it restricts model variations to a parametric family (or more formally the topological category proposed by Raghothama and Shapiro [1,17,27]). To improve its reusability, many resort to appropriate modeling guidelines that designers need to follow [7]. Typical ones include horizontal modeling [28], explicit reference modeling [29], and resilient modeling [30]. The idea is to reduce the use of feature dependencies such that a parametric model's rigidity can be loosened. Using those guidelines, larger model variation spaces are attained, but these spaces still fall into the category of parametric families. This limits their applicability to a structural family of models.

Another widely recognized solid modeling paradigm is the recent direct modeling, which attains model variations through direct manipulation of the model's boundary representation [31–34]. In principle, direct modeling can change a solid model to any shape. This, however, comes at a high price: feature information is lost, and parametric modification is no longer supported. Hence, it runs counter to the aim of retaining parametric capability in handling a structural family of models. Although there exist several methods to integrating direct modeling with parametrics, e.g., [35–37], they are still under development and not ready for practical usage.

**2.2. A posteriori variational design**

Posteriori means that a design's variations are not quite known in advance but are informed by product customization requirements. A common practice for handling this task is: when a new customization order comes in, the designer first searches for a base model, on the internet or in a proprietary model library, then modifies the model to repurpose it into a new one. Methods related to this routine can be classified into two levels: complete model retrieval and partial model recognition. The former refers to picking out a model from a collection of models, which should be close to the intended new design [38]. The latter recognizes local features that frequently occur in a CAD library [16].

For complete model retrieval, there are two basic types of approaches: (1) feature-based techniques and (2) shape-based techniques. The essential task here is defining a metric measuring model similarity. Feature-based techniques make use of high-level feature semantics to define the metric, while shape-based techniques rely only on low-level geometric information.

Commonly used feature semantics include feature types and feature dependency graphs [39–42]. Shape-based methods focus primarily on point clouds and mesh models. They usually conduct a direct geometric comparison between models, augmented with shape descriptors such as curvatures [43,44]. Very recently, deep learning are being introduced to generate data-driven feature semantics and shape descriptors, which have shown promising results [45,46]. It is undeniable that this body of methods are applicable to the problem considered in this work. However, they still require considerable manual input. For example, the user needs to construct a query model or sketch for those methods to start the searching. Also, those methods often output multiple candidate models rather than a single one, and consequently the designer still needs to do manual comparisons among models.

Partial model recognition is similar to conventional feature recognition but differs in that it aims to find frequently occurring features in the model library instead of identifying specific features on a single model. In the literature, this task is usually referred to as common design structure recognition [16]. (Other terms, although not very common, have also been used, such as design patterns [38] and subparts [47]). To extract the design structures, frequency-based approaches were used to cluster similar geometric entities and to pack them as features [15,16,38,47,48]. Unfortunately, those methods focus primarily on simple, detail features (e.g., holes and slots), not a master model for an entire model library.

The above review suggests that many researchers have investigated the variational design problem, and their methods are very inspiring to this work. Nevertheless, there is currently no method designated to variational design for a structural family of models. In the following, we present a master model-based variational design method. The master model not only can capture the structural pattern underlying the family but also can retain all features in the family. Using it, variational design for a structural family of models reduces to simple operations of removing unwanted detail features and adding new features, without the need for explicitly referring to the model collection.

## 3. Method overview and relevant notions

Fig. 1 shows the workflow of the proposed master model-based variational design, as well as its comparison with conventional variational design. As can be seen from the figure, conventional variational design requires several manual inputs, some of which are time-consuming. In particular, the designer needs to construct a query model that must be fine enough for a model retrieval algorithm to narrow all candidate models down to a single base model. The designer also needs to go through all models in the dataset to collect detail features that can be reused by the new design. Adding them to the base model is also a tedious process since feature dependencies are hard to manage manually. By contrast, the proposed variational design scheme constructs automatically a master model that can act like the refined base model in conventional variational design. What remains to be done by the designer is then to remove unwanted detail features, add new detail features (corresponding to new detail functions), and change parameter values, as shown in Fig. 2. All of these operations are lightweight.

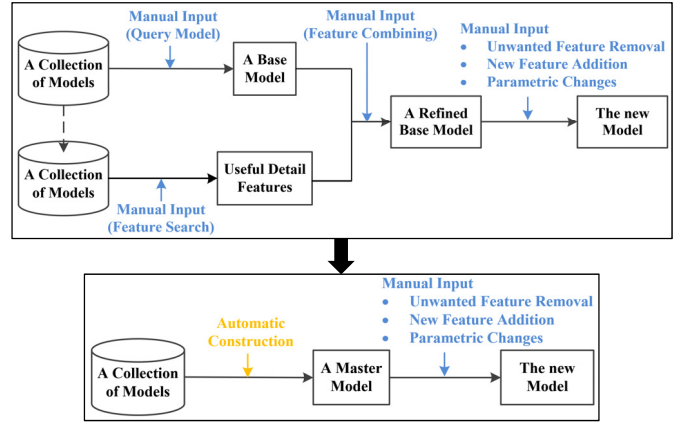

Figure 1. Workflow diagrams of master model-based variational design (Bottom) and conventional variational design (Up).

Clearly, the essential part of the proposed variational design scheme lies in automatic generation of the master model from a given collection of models. This paper restricts the model collection to be a structural family. *By structural we mean there is an underlying, shared structural pattern in the family's constituent models, and their detail features can vary as responses to product customizations.* Model collections of this sort are often not obtained through searching and saving models from publicly available libraries (e.g., the GrabCAD website), but are results of accumulating customized models of a specific product type within a company, e.g., the engine rotor discs designed by the Siemens energy department [49]. Due to this proprietary nature, the following three major characteristics were observed for a structural family of models:

(1) Models belonging to a same structural family have similar main features since main features dominate a model's rough shape and main function, refer to Fig. 2 for examples;
(2) The construction histories (feature generation and association) of those main features in individual models are similar because they are very likely modeled by a same team in a company; and
(3) Dimensions of individual models may vary significantly, so do their detail features.

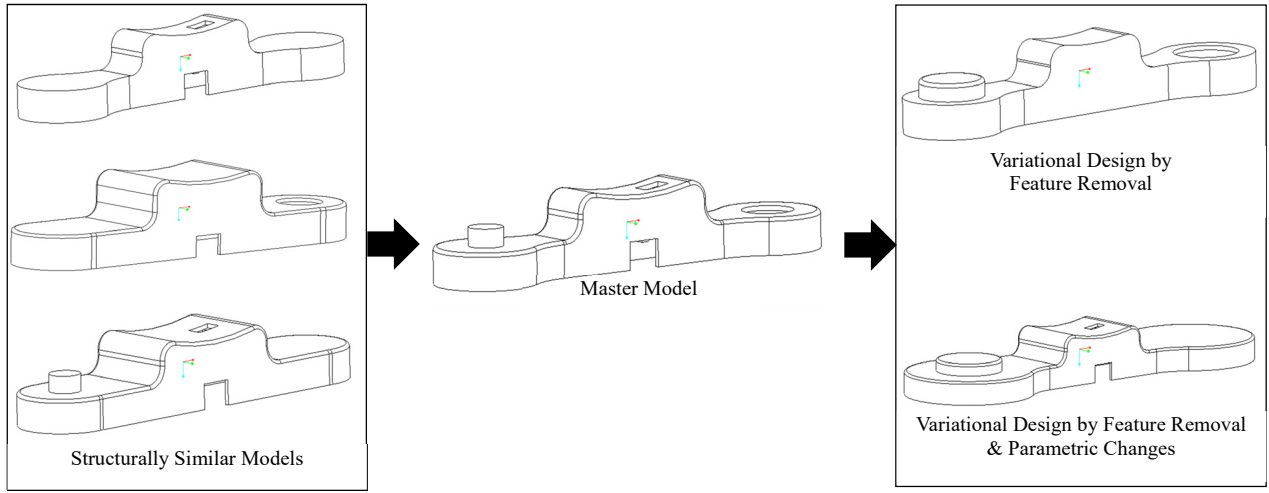

Figure 2. Illustration of master model-based variational design.

Our master model generation method will make use of the above characteristics. Following Characteristics (1) and (2), this work expresses the structural pattern underlying a model family in terms of the maximum common features (together with their associativity) among the family's members. It should, however, be noted that the maximum common features will cover a detail feature if it consistently occurs in all models. This over-coverage is generally acceptable because this kind of detail features must correspond to some critical functions, otherwise they cannot pertain over the course of all product customizations.

Characteristic (3) implies that a main feature, although being shared, could take various forms in individual models due to variations in dimensions and interactions with detail features. Determining which two instance features are the same is thus a critical task in master model generation. Another challenge lies in maximizing the common feature set. In the following sections, new/improved methods are to be presented to solve these challenges.

## 4. Master model generation

### 4.1. The proposed rolling snowball method

To extract the structural submodel composed of the maximum common features of a family, a simple iterative idea may be used. Fig. 3a outlines its basic steps, which iteratively identify the common feature set between a model in the family and the common feature set obtained so far. The resulting common feature set is maximal because any common feature will not be missed by the steps. This basic algorithm, however, has an important issue. Take the situation in Fig. 3b as an example. The algorithm will output a shared structural submodel composed of $A$ and $B$. Adding the left detail features to this submodel causes feature duplications, as indicated by the red ellipse in Fig. 3b. This issue is due to the possible commonality among detail submodels (composed of the left detail features, see Fig. 3b). Resolving the duplications is not trivial because it requires an exhaustive extraction of all common features among any subset of detail submodels.

In view of the above issue, this paper proposes a new method, as outlined in Fig. 4a. It basically moves from the "intersection-like" idea described above to a "union-like" idea. Specifically, after finding the common feature set between two models, we immediately merge them into one, as an intermediate master model. Starting from the first two models in the family and iterating over the other models one-by-one, the find-then-merge procedure will accumulate to the intended master model. The method's cumulative process is similar to rolling snowball, which explains its name.

Feature duplications can be automatically handled by the merge steps, as show in Fig. 4b. More importantly, this method is *order-independent*, which means that a same master model will be generated regardless of the model sequence used. This property is very important to repeatability of the proposed method. To show this property, we again use the example in Fig. 4b. There are three model sequences for the example models: $(i,j,k)$, $(i,k,j)$, and $(j,k,i)$. Fig. 4b and Fig. 5 show the results of applying the proposed method to them. A same master model has been successfully generated. (It should, however, be noted that the final maximum common feature set may be different, as shown by the results in Figs. 4b and 5. Some sequences may even give incorrect results on maximum common features, as shown in Fig. 5 (where the correct result should be composed merely of $A$ and $B$). This problem is trivial and can

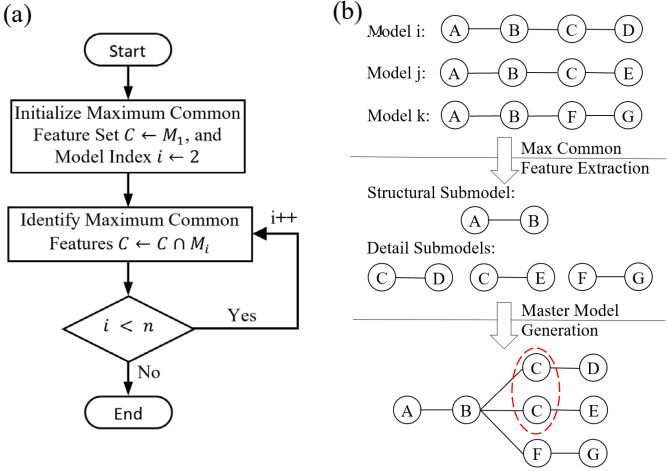

Figure 3. Iterative maximum feature set extraction method (a) and illustration of feature duplication issues (b).

be easily solved by keeping the smallest common feature identification result over the course of iterations.)

Having outlined the proposed method (Fig. 4a), the last missing piece in master model generation is to identify the maximum common feature set between two given models. The next subsection is devoted to this task.

*4.2. The maximum feature set identification algorithm*

Following existing work on this topic, the sub-graph isomorphism strategy [16] is chosen to extract the maximum common features between two models. A direct implementation of this strategy is, however, found to be slow for problems with repeated binary model comparisons, especially for our cases where one model's size gets larger and larger as the algorithm goes. To accelerate sub-graph isomorphism, a modified version of it is to be used, which consists of two consecutive steps: (1) a rough feature match based solely on features; and then (2) a fine feature match based on additional feature dependencies. The former step is local (i.e., individual features) and focuses on finding candidate feature pairs that can be matched. It can help avoid unnecessary feature match attempts in sub-graph isomorphism. The latter step is global (i.e., collective features) and makes use of attributed feature dependency graphs to further reduce the candidate feature pairs to one-to-one feature matches.

*4.2.1. Rough feature matching*

Instead of doing an exhaustive check for all feature pairs of two given models, a top-down subdivision method is used to group feature pairs. This method can reduce the computational complexity from $O(mn)$ to $O(m+n)$, where $m, n$ denote the number of features in two models. The idea is to iterate over all features of two models and distribute them into a number of buckets, each of which represents a feature class according to a feature comparison criterion.

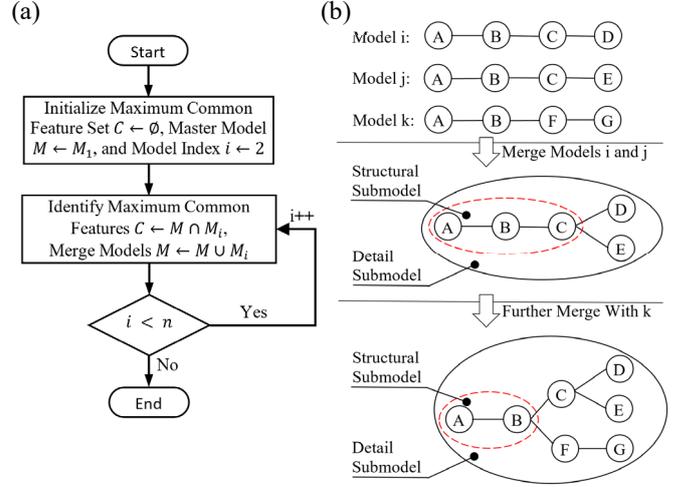

Figure 4. The rolling snowball method (a) and an example of its execution (b). Nodes labeled as A-G represent features and edges describe feature dependencies.

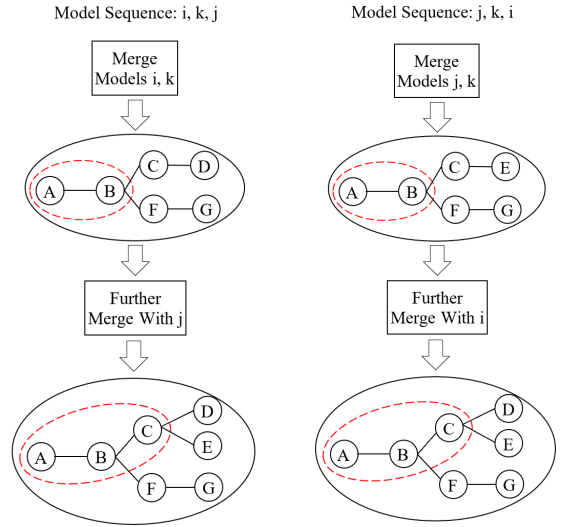

Figure 5. Examples of the order-independent property.

Algorithm 1 shows the detailed procedures used. The algorithm first generates a set of buckets (Line 1). Each bucket is denoted as $b_i$ and divided into two separate sub-buckets $b_i^1, b_i^2$ to hold features from the two models, respectively. It then checks each feature against the criterion to see which bucket it belongs to (Line 3). Here hash maps can be used to generate the bucket indices. After getting the corresponding bucket, it simply adds the feature under checking to the corresponding sub-bucket (Lines 4-8). Repeat the above procedures until all features are checked. Each bucket finally contains two groups of features, and any two of their respective elements form a candidate feature pair.

---

**Algorithm 1**: Rough Feature Matching

**Input:** $M_1, M_2, \varphi$ — The two models to be compared, and the comparison criterion

**Output:** $B$ — Buckets containing candidate feature pairs

1. $B \leftarrow \{b_1, \cdots, b_n\},\ b_i = \{b_i^1, b_i^2\} = \emptyset, i = 1, \cdots, n$
2. **for** each feature $f_i$ in $(M_1, M_2)$ **do**
3. $\quad idx \leftarrow getBucketIndex(\varphi(f_i))$
4. $\quad$ **if** isFromFristModel$(f_i) = TRUE$ **then**
5. $\quad\quad b_{idx}^1 \leftarrow b_{idx}^1 \cup f_i$
6. $\quad$ **else**
7. $\quad\quad b_{idx}^2 \leftarrow b_{idx}^2 \cup f_i$
8. $\quad$ **end if**
9. **end for**
10. **Return** $B$

---

The above description is for a single comparison criterion. In practice, multiple comparison criteria are needed to achieve satisfactory results. A hierarchical version of Algorithm 1 is used to handle multiple comparison criteria. Let the criteria be represented by $(\varphi_1, \cdots, \varphi_l)$. For the (i)-th criterion, the algorithm accepts as input the outputs of the (i-1)-th criterion, i.e., a set of buckets $\{b_1, \cdots, b_n\}$. It then applies Algorithm 1 to each bucket $b_i$, or more specifically the two sub-buckets of $b_i$, rather than the original models $M_1, M_2$. The outputs of the (i)-th criterion are, again, used as inputs for the (i+1)-th criterion. In this way, the original models $M_1, M_2$ are subdivided into buckets by the criterion $\varphi_1$; the following criteria $\varphi_1 - \varphi_m$ continue to subdivide those buckets into smaller buckets.

The feature comparison criteria used in this work are as follows:
(1) **Feature Type.** The basic feature types defined in CAD systems such as datum, protrusion, and cut.
(2) **Feature Sub-type.** Feature types defined to help construct user-defined features, including extrude, revolve, and loft.
(3) **Attributed Face Graph.** Face adjacency graph of a feature, together with face types, face genus, and face-face concavity/convexity.
(4) **Sketch Topology.** Loop numbers and their closeness of 2D sketches. We divide sketch topologies into four types: closed loop, single non-closed loop, multi-closed loop, and multi-non-closed loop.
(5) **Sketch Type.** Edge types and quantities of sketches. The edge type describes if an edge is linear or curved.

*4.2.2. Fine feature matching*

At this step, feature dependencies are taken into consideration to refine the candidate feature pairs, eventually leading to the maximum common feature set between two models. To do so, a combination of attributed feature graph representation and sub-graph isomorphism is used.

Attributed graph is not new and has been widely used in CAD, especially for feature recognition [50]. It is defined as an ordinary graph with additional descriptions of the graph's nodes and/or edges. Let a graph be represented by $G = (V, E)$, where $V$ denotes its vertices, and $E$ its edges linking the vertices. As feature dependencies are directed, a graph representing a feature model is directed (i.e., edges in $E$ are ordered). An attributed graph can be represented as $G = (V, E, D_v, D_E)$, where $D_v$ denote feature descriptors (i.e., feature attributes), and $D_E$ descriptors about feature dependencies.

In this work, only feature descriptors will be used, and thus the graph reduces to $G = (V, E, D_v)$. Specifically, our feature descriptors are based on the bucketing results of the above rough matching. Let those buckets be denoted by $\{\{b_1^1, b_1^2\}, \cdots, \{b_n^1, b_n^2\}\}$. We assign a same code to features belonging to a bucket and different codes for features in different buckets. For example, each feature can be assigned a code same as the subscripts of the bucket indices. These simple codes are then used as feature descriptors to characterize feature commonality in constructing the attributed graphs for two models undergoing comparison.

With two attributed graphs in place, existing graph algorithms can be readily applied to finding the common sub-graph between them. The authors have previously presented an improved McGregor method to identify common sub-graphs [51]. The method is applicable to the attributed graphs in this work. It is thus the sub-graph isomorphism algorithm of choice. The basic idea of the McGregor method is to use depth-first search to match subgraphs. The improvements made are briefly summarized below:
(1) **Model Simplification.** Removing detail features, e.g., blending and small holes.
(2) **Depth Limitation.** Setting the maximum depth of tracing back in depth-first search. McGregor suggests the tracing back depth be the same as the number of graph nodes. For a feature model, its dependencies are usually hierarchical, and thus the depth can be limited while keeping the outcome unchanged.

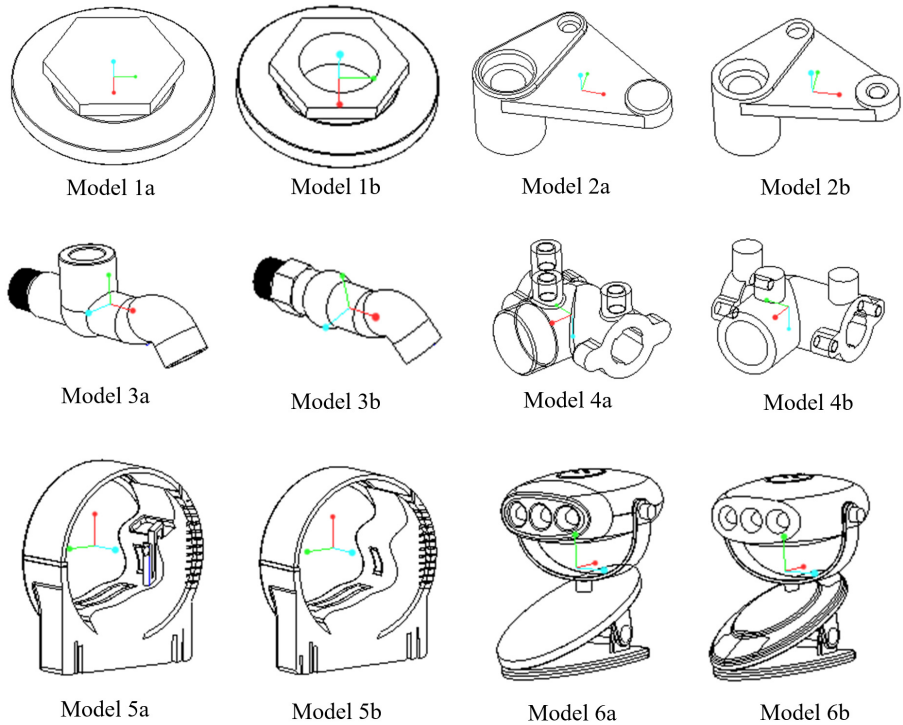

Figure 6. Models for efficiency comparisons.

At this point, we can assemble the attributed graph construction algorithm and the improved McGregor algorithm stated above with Algorithm 1 to form a maximum common feature set extraction algorithm, which in turn can be input into the rolling snowball method outlined in Fig. 4a to form a complete master model generation algorithm for a structural family of CAD models.

## 5. Results and discussion

The previously presented methods have been implemented using the PTC Creo software package and its APIs. Three case studies and six comparisons are to be presented to demonstrate the effectiveness of the proposed method. They were chosen carefully to have increasing model complexity. For example, Case study 1 considered a model composed primarily of planar surfaces; Case study 2 analyzed a model containing more features and surface types; Case study 3 involved a model with many features and complex feature dependencies.

### 5.1. Case studies and comparisons

The comparisons were made between the proposed algorithm described in Section 4.2 and a direct implementation of the sub-graph isomorphism strategy. Fig. 6 shows the six pairs of models used to carry out the comparisons, and Table 1 summarizes the statistics (on a 2.6 GHz Intel Core with 8G memory). In the second and third columns (i.e., "# of Nodes" and "# of Edges"), the numbers are the average features and dependencies of two models under comparison. The fifth column corresponds to test results of the direct implementation version, and the sixth is for the proposed method.

Case study 1 considered a family of five models, as shown in Figs. 7a-7e. This family represents a relatively simple example of master model generation since the number of features are small, the surface types are primarily planar, and particularly the variations of detail features of the models are quite obvious. The generated master model for this family is shown in Fig. 7f. Some important detail features on the master model have been highlighted using red rectangles.

Case study 2 involved a family of more complex models (Figs. 8a-8f). The number of features involved in those models doubles those in Case study 1. The feature dependencies are also more complex than the previous models. For this family, the proposed method was also able to successfully generate a master model (Fig. 8g), within 24 seconds.

Case study 3 analyzed an even more complex family of models, as shown in Figs. 9a-9g. These models have an average number of 99 features, their feature variations are considerably large. As can be seen from the master model in Fig. 9h, all detail features have been correctly added to the master model. The total time used in generating this master model is 78.44 seconds.

Table 1. Comparison results of the proposed method with the direct method, using the models in Fig. 5.

| Models | Attributed Graph Complexity | | | Computation Time (s) | | |
| --- | --- | --- | --- | --- | --- | --- |
| | # of Nodes | # of Edges | # of Common Features | Original | Accelerated | Improvement |
| Models 1 | 12 | 16 | 10 | 0.31 | 0.009 | ×34 |
| Models 2 | 14 | 27 | 10 | 3.54 | 0.013 | ×278 |
| Models 3 | 26 | 41 | 21 | 4.86 | 0.015 | ×320 |
| Models 4 | 30 | 56 | 22 | 19.04 | 0.014 | ×1386 |
| Models 5 | 82 | 194 | 61 | 376.83 | 0.206 | ×1830 |
| Models 6 | 91 | 235 | 76 | 686.44 | 0.220 | ×3118 |

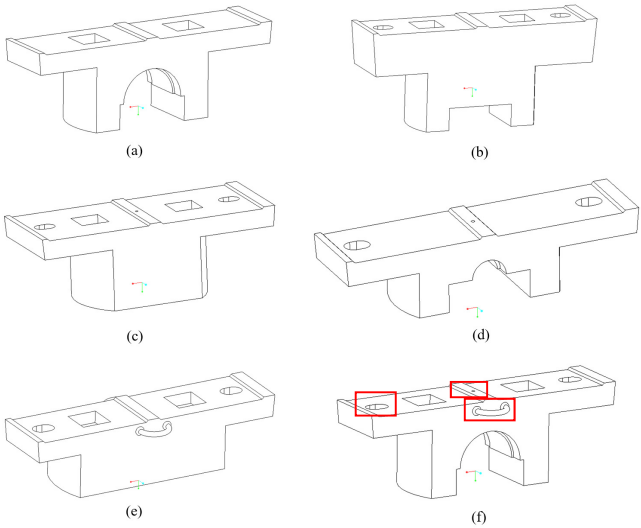

Figure 7. Case study 1: master model generation (f) for a family of models (a)-(e). Red rectangles indicate some typical common features generated.

Based on the master model generated in Case study 3, we carried out several variational design examples in the PTC Creo modeling environment, as shown in Fig. 10. Fig. 10a displays the generated master model in PTC Creo, and its feature information is listed on left Model Tree panel. To modify this model, one simply needs to suppress unwanted detail features, edit parameter values, or add new detail features. Figs. 10b-10d show three such editing examples, where the blue rectangles indicate removed features and red rectangles highlight parameter changes.

*5.2. Discussion and limitations*

From the comparisons shown in Fig. 6 and Table 1, the proposed algorithm is seen to accelerate the original method significantly. Large improvements are expected and confirmed for complex models such as Models 5 and 6. Even for simple models which leave a small room for improvement, the proposed method is still able to accelerate the process by a notable percentage, above 30 times. In addition, there is a positive relationship between model complexity and efficiency gain, as shown in Fig. 11.

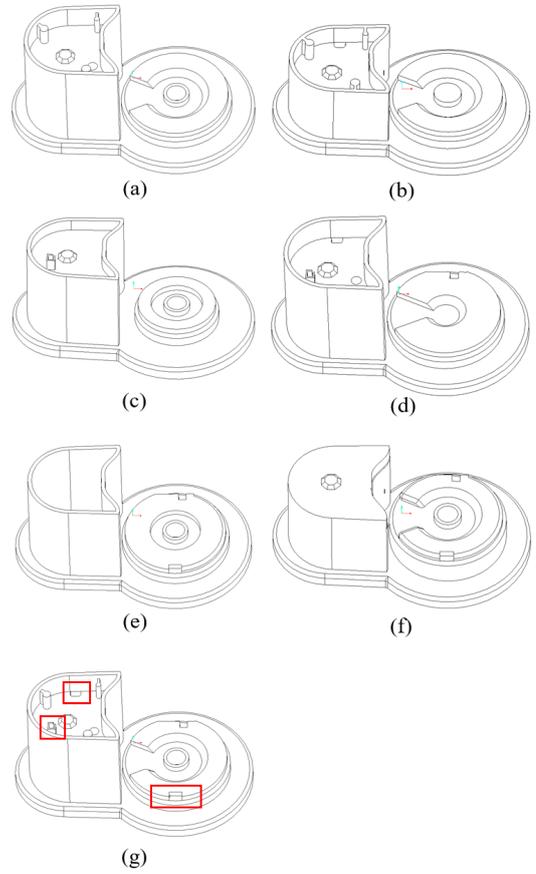

Figure 8. Case study 2: master model generation (g) for a family of models (a)-(f). Red rectangles indicate some typical common features generated.

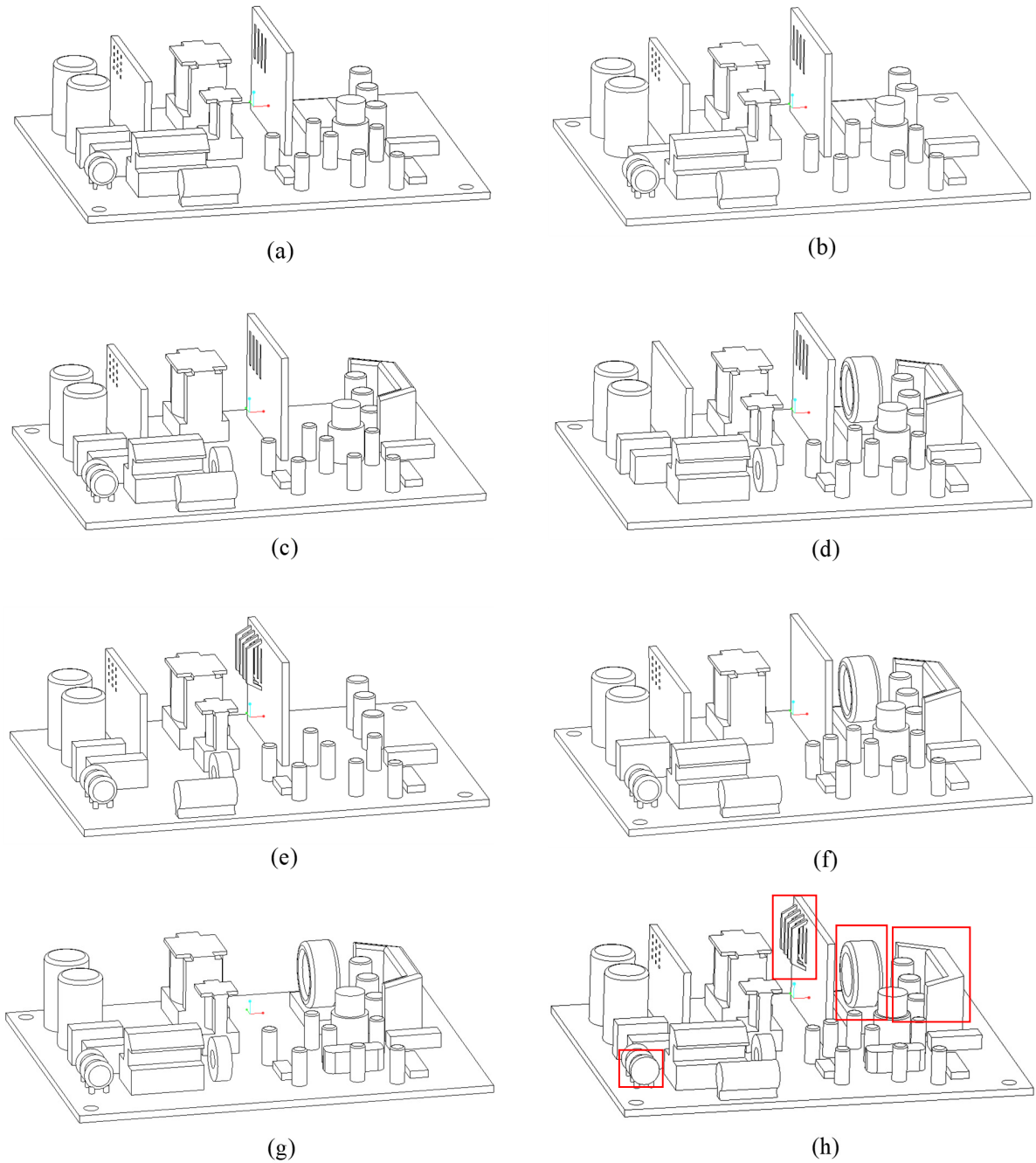

Figure 9. Case study 3: master model generation (h) for a family of models (a)-(g). Red rectangles indicate some typical common features generated.

The case studies shown in Figs. 7-9 demonstrate the effectiveness of the proposed rolling snowball method. For each case study, the proposed method has successfully generated its master model. As shown in Figs. 7f, 8g and 9h, no feature duplications occurred, and all features resided at proper places on the master models. The feature history shown in the left side of Fig. 10a also confirms that feature dependencies were managed well.

Despite the effectiveness and efficiency demonstrated, it should be noted that there are parameters that need to be tuned to achieve such good results. When conducting the comparisons shown in Fig. 6, the authors found that the depth limitation parameter used to limit the maximum depth of tracing back in the sub-graph isomorphism method (Section 4.2.2) plays an important role in the overall computation time of the method. If the depth is small, e.g., less than 3, the algorithm may be trapped in local minimums. If the depth is large, say more than 15, enormous trace back furcations will be tried by the algorithm, which eventually results in high computation time. To completely solve this issue, an automatic mechanism to set the depth parameter is needed. However, such a mechanism still remains unknown, and further development is required. Fortunately, from our experience, a depth from 5 to 9 can balance properly the optimums and running time.

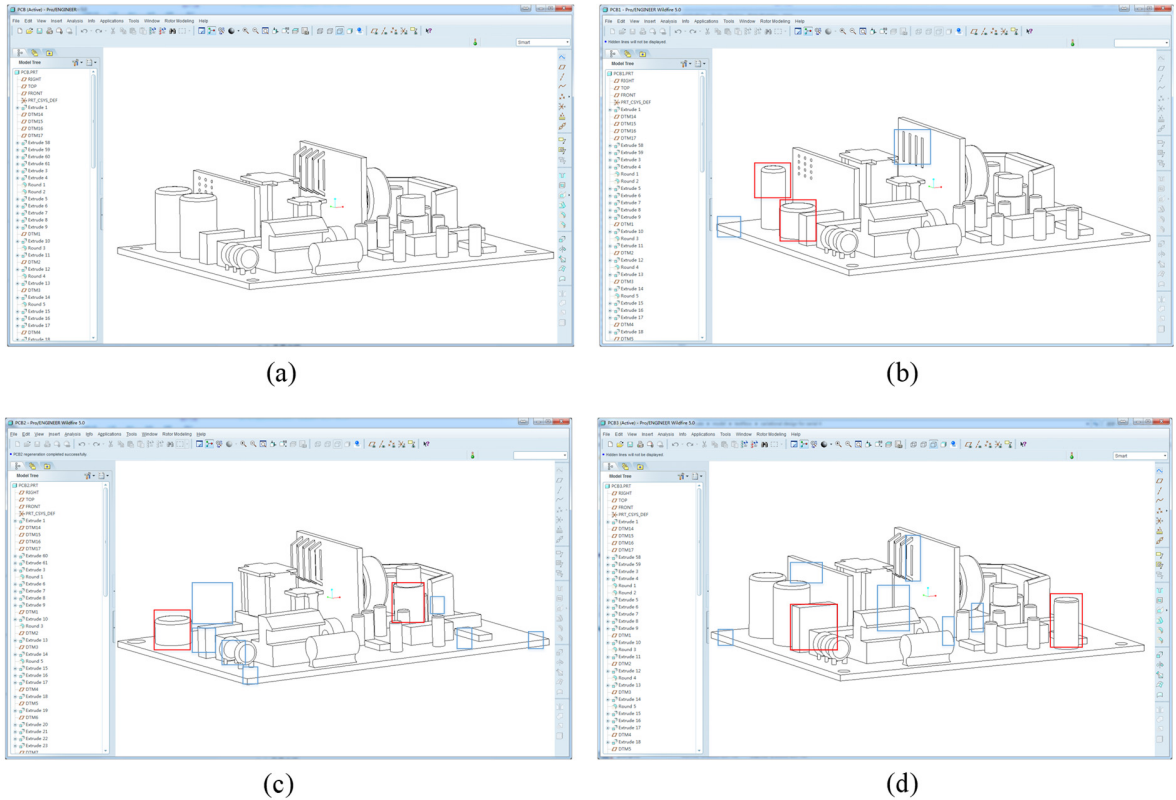

Figure 10. Variational design examples based on the master model from Case study 3.

## 6. Conclusion

A new method has been presented in this paper to facilitate variation design for a structural family of CAD models. The basic idea behind this method is to generate a master model that works as a representative model for the family and contains all of its features. With this master model, variational design reduces to several easy and fast modeling operations. These advantages are essentially achieved by (1) formulating the underlying structural pattern of models as a submodel composed of the maximum common features, and (2) casting the submodel generation problem as a succession of binary model comparison problems, i.e., the rolling snowball method. New/improved algorithms have been presented to implement this method, and a series of case studies and comparisons have been conducted to validate the method.

Although the presented method is seen to be quite effective in the case studies conducted, there are a few limitations (and therefore future work) that should be noted here. The proposed method, in its current form, requires that model families input into the proposed method are strictly structural families. As a result, the method's applicability is limited. One way to relax the requirement is to include model families where the majority of models are structurally similar. This gives rise to the need for a method that can automatically identify and remove models that are not structurally similar to the majority. It is among the research studies to be carried out by the authors.

Another interesting improvement direction is to have an automatic method to set the depth limit parameter in sub-graph isomorphism. The present work has identified a range, 5-9, for the proposed method to work properly. It would, however, be better to have an adaptive method to set the parameter according to the complexity of the specific model considered. In previous experiments, the authors have observed certain correlation between model complexity and depth limit. Clarifying this correlation and then generating a systematic set of rules accordingly are very promising.

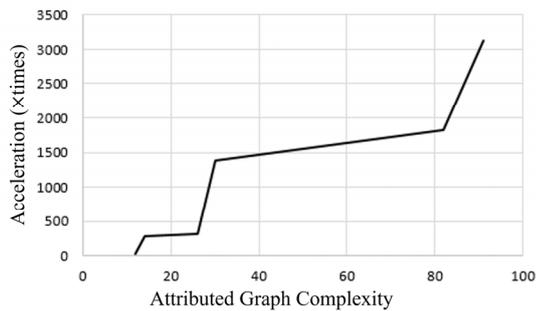

Figure 11. Analysis of the relationship between complexity and acceleration.

**Acknowledgments**

The authors are very grateful to the financial support from National Natural Science Foundation of China (#62102355) and NSF of Zhejiang (# LQ22F020012).